\documentclass[12pt,a4paper]{article}
\usepackage{graphicx}
\usepackage{jcappub}
\usepackage{epsfig} 
\usepackage{amsmath}
\usepackage{amsfonts}
\usepackage{graphicx}
\usepackage{amssymb}

\title{An Extended Zel'dovich Model for the Halo Mass Function}
\author{Seunghwan Lim}
\author{$\&$ Jounghun Lee}
\affiliation{Astronomy Program, FPRD, Department of Physics and Astronomy,
Seoul National University, Seoul 151-747, Korea}
\emailAdd{shlim@astro.snu.ac.kr,  jounghun@astro.snu.ac.kr}

\abstract{ 
A new  way to construct a fitting formula for the halo mass function is presented.  
Our formula is expressed as a solution to the modified Jedamzik matrix equation 
that automatically satisfies the normalization constraint.  The characteristic parameters 
expressed in terms of the linear shear eigenvalues are empirically determined by fitting 
the analytic formula to the numerical results from the high-resolution N-body simulation 
and found to be independent of scale, redshift and background cosmology.  
Our fitting formula with the best-fit parameters is shown to 
work excellently in the wide mass-range at various redshifts: The ratio of the analytic 
formula to the N-body results departs from unity by up to $10\%$ and $5\%$ over 
$10^{11}\le M/(h^{-1}M_{\odot})\le 5\times 10^{15}$ at $z=0,\ 0.5$ and $1$ for 
the FoF-halo and SO-halo cases, respectively. }
\begin{document}
\maketitle

\section{Introduction}\label{sec:intro}

The era of precision cosmology that reigned supreme in the last century 
has waned, rendering its glory to the advent of a new era of accurate 
cosmology \cite{peebles02}. 
While the goal of the previous era was the precise measurements 
of a model-dependent finite set of the cosmological parameters, the 
accurate cosmology aims at addressing more fundamental
issues such as what the origin of cosmic inflation is, what 
the limitation of the general relativity is, what drove the cosmic 
acceleration and etc, which requires to explore almost infinite parameter 
space spanned by all possible cosmological and physical scenarios.

In precision era, resorting solely to the high-resolution simulations 
suffices to make precise theoretical predictions for the determination of 
the cosmological parameters. However, in accuracy era, given that it is 
extremely expensive and inefficient to sweep through the infinite parameter 
space using the simulations alone, developing robust analytical 
guidelines is highly desired to complement the numerical experiments.
The mass function of bound halos has been spotlighted as one of those few 
cosmological probes which allow an analytical approach.

It was ref.~\cite[][hereafter PS]{PS74} who developed for the first time an analytic formalism for the 
halo mass function under several simplified assumptions on the halo formation process. Later, 
ref.~\cite{bond-etal91} introduced the excursion set theory  to  analytically derive the mass function 
from physical principles  and found that the excursion set theory yields the same PS mass 
function for the special case of the sharp-$k$ space filter \cite[see also][]{PH90}.  
In the light of these landmark works, rapid progress has been made to improve the original 
excursion set formalism in two different directions.  One direction was to refine the analytical 
prescriptions by incorporating into the excursion set theory more realistic  
aspects of the halo formation process such as the occurrence of the clouds-in-clouds, 
spatial correlations among the proto-halos, tidal effect from the surrounding matter distribution, 
diffusive nature of the density threshold, and etc \cite[e.g.,][]{jedam95,yano-etal96,audit-etal97,monaco97a,monaco97b,SMT01,CL01,MR10a,MR10b,MR10c,CA11a,CA11b,PLS12,MS12,PS12}.

The other direction focused on finding a more accurate analytic formula for the halo mass function by modifying 
the functional form of the excursion set mass function 
\cite{LS98,ST99,jenkins-etal01,reed-etal03,warren-etal06,tinker-etal08,mice10,pph10}. 
The modified functional forms  are characterized by free parameters whose best-fit values 
have to be empirically determined by fitting the formulae to the high-resolution N-body results.  Although the 
recent analytic formulae  have been found to agree excellently with the N-body results,  
the best-fit values of their characteristic parameters turned out to likely vary with the  mass 
scale, redshift and background cosmology  \cite[e.g.,][]{mice10,lee12}.
  
In this paper, we present a new fitting formula for the halo mass function which is characterized by the 
free parameters independent of scale, redshift and background cosmology.  
The Jedamzik formalism \cite{jedam95} is adopted as our statistical framework within which the halo 
mass function is automatically normalized. 
Throughout this paper, we assume a flat $\Lambda$CDM cosmology with the WMAP7 parameters 
\cite{wmap7} (unless stated otherwise) and use the CAMB code \cite{camb} for the evaluation of the 
$\Lambda$CDM linear power spectrum.

\section{Analytic framework: a review }
\label{sec:review}

Most of the previous formulae for the halo mass functions were constructed in the the following 
analytic framework \cite{LS98,ST99,jenkins-etal01,reed-etal03,warren-etal06,tinker-etal08,pph10,mice10}: 
\begin{equation}
\label{eqn:frame}
\frac{dN(M, z)}{d\ln M}=\frac{\bar{\rho}}{M}\frac{d\ln\sigma^{-1}}
{d\ln M}f[\sigma(M,z)].
\end{equation}
Here $dN(M, z)/d\ln M$ is the differential number density of the bound halos in a logarithmic 
mass interval of $[\ln M,\ \ln M +d\ln M]$ at redshift $z$ per unit volume,  
$\sigma(M,z)\equiv b(z)\sigma(M,0)$ is the rms fluctuation of the linear density field 
smoothed on the mass scale $M$ at redshift $z$ where $b(z)$ 
is the linear growth factor satisfying the condition of $b(0)=1$, $\bar{\rho}$ is the mean 
mass density of the Universe, and $f(\sigma)$, called the multiplicity function, 
represents a differential volume fraction occupied by those regions which 
satisfy a prescribed collapse condition for the halo formation. 
In jargon of the excursion set theory, a random walk proceeds as the time-like 
variable $\sigma$ increases until it hits a specified collapse barrier.  The number of 
those random walks which first cross a given collapse barrier on scale of $\sigma(M,z)$ 
is proportional to the number of the bound halos of mass $M$ formed at $z$ \cite{bond-etal91}.  

The functional form of $f(\sigma)$ that is the key quantity in equation (\ref{eqn:frame}) 
depends on  the correlations among random walks as well as on the shape of the collapse barrier. 
The latter is determined by the underlying dynamics while the former is related to the shape of 
the filter used to smooth the linear density field.  As mentioned in section \ref{sec:intro}, 
ref.~\cite{bond-etal91} showed that the original PS mass function can be derived from the excursion set 
formalism under the assumptions that the random walks are Markovian (i.e., uncorrelated) 
corresponding to the sharp-$k$ space filter and that the collapse barrier has a flat shape, 
$\delta=\delta_{sc}$, corresponding to the spherical dynamics. The PS multiplicity 
function derived from the excursion set theory is written as
\begin{equation}
\label{eqn:ps}
 f_{\rm PS}[\sigma(M,z)] = \sqrt{\frac{2}{\pi}}
\frac{\delta_{sc}}{\sigma(M,z)}
\exp\left[-\frac{\delta_{sc}^2}{2\sigma^2(M,z)}\right],
\end{equation}
where the height of the spherical collapse barrier $\delta_{sc}\simeq 1.686$ depends only very weakly on the 
background cosmology \cite{eke-etal96}.  

As also mentioned in section \ref{sec:intro}, much effort has been made to find a better fitting formula by 
phenomenologically modifying the above PS multiplicity function. For instance, ref.~\cite{tinker-etal08} 
modified the PS multiplicity function into 
\begin{equation}
\label{eqn:tinker}
 f_{\rm tinker}(\sigma) = A\left[\left(\frac{\sigma}{b}\right)^{-a} +1\right]
\exp\left(-\frac{c}{\sigma^{2}}\right),
\end{equation}
where the four coefficients, $A$, $a$, $b$ and $c$, are  free parameters whose best-fit values were determined 
empirically by comparing equation (\ref{eqn:tinker}) with the N-body results.  Their formula were found to 
agree with the high-resolution N-body results with accuracy up to maximum $10\%$ error for the case that a 
bound halo is identified by the friends-of-friends (FoF) algorithm \cite{davis-etal85}.

For the case that a bound halo is identified by the spherical overdensity algorithm \cite{LC94}, 
the following formula for the multiplicity function provided by ref.~\cite{pph10} has 
been found to work better \cite[see also][]{jenkins-etal01,warren-etal06}:
\begin{equation}
\label{eqn:pph}
 f_{\rm pillepich}(\sigma) = \left[D+B\left(\frac{1}{\sigma}\right)^{A} \right]
\exp\left(-\frac{C}{\sigma^{2}}\right).
\end{equation}
The best-fit values of the four coefficients $A$, $B$, $C$ and $D$ were determined by comparing 
equation (\ref{eqn:pph}) to the results from the high-resolution N-body simulations for a $\Lambda$CDM 
cosmology with the WMAP5 parameters \cite{wmap5}. Their formula was demonstrated to reach the accuracy 
level up to maximum $5\%$ error over a wide mass range of 
$2.4\times 10^{10}\le M/(h^{-1}M_{\odot})\le 10^{15}$. For the other fitting formulae formulated within the 
framework of equation (\ref{eqn:frame}), see Table 7 in ref.~\cite{pph10}. 

It is worth noting that the multiplicity function in equation  (\ref{eqn:frame}) does not  
automatically satisfy the normalization constraint of $\int_{0}^{\infty} d\sigma f(\sigma)=1$ (under the 
assumption that all initial regions would eventually collapse to form bound halos) but its overall 
amplitude had to be treated as an additional fitting parameter. For instance, the parameter $A$ in equation 
(\ref{eqn:tinker}) and $D$ in equation (\ref{eqn:pph}) had to be introduced to satisfy the normalization 
constraint. 

It has been conventionally thought that the normalization of the halo mass function is related to the issue of 
taking into account the occurrence of the clouds-in-clouds (underdense regions embedded in larger 
overdense regions) and that for the special case of Markovian random walks with flat spherical collapse 
barrier the overall normalization factor of $2$ introduced by PS can be justified by 
the excursion set theory \cite{PH90,bond-etal91,jedam95} as a solution to the clouds-in-clouds problem. 
Very recently, however, ref.~\cite{PLS12} argued that the normalization of the halo mass function in fact 
reflects the limitation of the excursion set approach itself having strong dependence on the correlations 
of random walks. 

Given the downside of equation (\ref{eqn:frame}) that it requires an additional free parameter to satisfy 
the normalization of the halo mass function,  here we instead consider the Jedamzik equation \cite{jedam95} 
as our analytic framework which automatically yields the normalized mass function: 
\begin{equation}
\label{eqn:jedam_sp}
F(M)=\int_{M}^{\infty}dM^{\prime}\frac{M^{\prime}}{{\bar\rho}}\frac{dN}{dM^{\prime}}
P(M, M^{\prime}; M^{\prime}\, {\rm being}\,{\rm isolated})\ ,
\end{equation}
where $F(M)$ is the cumulative volume fraction occupied by those random walks which exceed the collapse 
barrier on the mass scale $M$, while $P(M, M^{\prime}; M^{\prime}\, {\rm bing}\, {\rm isolated})$ is the 
conditional probability that a random walk exceeds the collapse barrier on the mass scale $M$ 
provided that it for the first time {\it just touched} the collapse barrier on some larger mass scale 
$M^{\prime}$. That is, {\it the isolated} $M^{\prime}$ means that the random walk has never touched (nor 
exceeded) the collapse barrier on mass scales larger than $M^{\prime}$.  

For the special case of the sharp-$k$ space filter (Markovian random walks), the conditional probability 
$P(M, M^{\prime}; M^{\prime}\, {\rm bing}\, {\rm isolated})$ can be simply given as the conditional probability  
$P(M, M^{\prime})$ that a random walk exceeds the collapse barrier on the mass scale $M$ provided it just 
touched the collapse barrier on some larger mass scale $M^{\prime}$ \cite{jedam95}.
For non-Markovian case (corresponding to the cases of the top-hat or Gaussian filters), however, we have 
inequality $P(M, M^{\prime}; M^{\prime}\, {\rm bing}\, {\rm isolated})\ne P(M, M^{\prime})$, and thus it is 
expected that $P(M, M^{\prime}; M^{\prime}\, {\rm bing}\, {\rm isolated})$ would have much more complicated 
expression. Nontheless, in the current work, we use $P(M, M^{\prime})$ for the conditional probability in 
equation (\ref{eqn:jedam_sp}), since our goal here is not to estalish a physical model but just to find a 
new fitting formula for the halo mass function which is otherwise devoid of physical content.

\section{Extension of the Zel'dovich model}
\label{sec:ezl}

The original Jedamzik formalism employed the spherical collapse dynamics for which the collapse 
barrier is flat (i.e., scale-independent), expressed in terms of the linear density contrast, just as in the original 
PS formalism. It has long been realized, however, that the condition for a given Lagrangian region to form a 
bound halo should depend not only on its initial spherically averaged overdensity but also on the tidal shears 
from the surrounding matter distribution \cite[e.g.,][]{monaco97a,monaco97b,audit-etal97,LS98,SMT01}. 
The tidal shears have an effect of disturbing the gravitational collapse and deviating the collapse process 
from spherical symmetry.   To take into account the tidal shear effect, the gravitational collapse process 
should be described by more realistic ellipsoidal dynamics rather than by the simplified spherical dynamics.
Unlike the spherical case, however, there is no unique collapse condition for the case of the ellipsoidal collapse 
process \cite{BM96}. Among several different ellipsoidal dynamics that were used in the literature, the Zel'dovich 
approximation \cite{zel70} is one of those models which can be relatively easily implemented into the mass 
function formalism since its ellipsoidal collapse condition is simply given in terms of the three eigenvalues, 
$\lambda_{1},\ \lambda_{2},\ \lambda_{3}$ 
(in a decreasing order, $\lambda_{1}\ge\lambda_{2}\ge\lambda_{3}$) of the initial shear tensors 
$(T_{ij})$.  

In several literatures, the Zel'dovich approximation has already been used as an underlying dynamics 
for the halo mass function. For example, refs.~\cite{monaco97a,monaco97b} assumed that the gravitational 
collapse occurs when the largest shear eigenvalue, $\lambda_{1}$, reaches some critical value, 
$\lambda_{1c}$, which corresponds to the collapse along the first principal axes of the initial shear tensors 
in the Zel'dovich approximation.  Ref. \cite{LS98} derived the halo mass function with ellipsoidal 
collapse barrier of $\lambda_{3}=\lambda_{3c}$ that corresponds to the third principal axis collapse 
\cite[see also][]{audit-etal97} according to the dynamical guideline of the Zel'dovich approximation.

Those preivous works which use the Zel'dovich approximation as a dynamical guideline assume that the 
gravitational collapse occurs when the shear eigenvalues are all positive. Unfortunately, however, 
ref.~\cite{porciani-etal02} have shown that one of the initial shear eigenvalues at a proto-halo region has 
different sign from the other two, which basically invalidates this fundamental assumption. Henceforth, 
those previous models based on the Zel'dovich approximation are not physical ones but only  mechanistic 
prescriptions to determine the fitting formula for the halo mass function.
Apart from the invalid collapse condition,  the previous works based on the Zel'dovich model also suffered 
from the normalization problem which is attributed to the fact that even for the case of the sharp-$k$ space filter 
$P(M, M^{\prime}; M^{\prime}\, {\rm bing}\, {\rm isolated})\ne P(M, M^{\prime})$ if the collapse conditions 
are expressed in terms of the shear eigenvalues. 
Very recently, ref.~\cite{LL12} have suggested that the normalization of the halo mass function in the 
Zel'dovich model can be satisfied if the Jedamzik formalism is used as a statistical framework.

Adopting a similar statistical strategy to that of ref.\cite{LL12}, we  construct an extended Zel'dovich model (EZL) 
for the halo mass function for three different cases:
For the one-dimensional case (1D EZL), the free parameter is given as the threshold of the 
smallest shear eigenvalue: ${\vec \lambda}_{c}=\{\lambda_{3c}\}$. 
For the two dimensional case (2D EZL), theere are two free parameters given as the thresholds of the second 
to the largest and the smallest eigenvalues:  ${\vec \lambda}_{c}=\{\lambda_{2c},\ \lambda_{3c}\}$, 
while for the three dimensional case (3D EZL), there are three free parameters given as the thresholds of 
all three shear eigenvalues: ${\vec \lambda}_{c}=\{\lambda_{1c},\ \lambda_{2c},\ \lambda_{3c}\}$. 
We emphasize here that the thresholds of the shear eigenvalues are not true physical collapse barriers 
but just free parameters that characterize the halo mass function in the mechanistic Zel'dovich model.

Expressing the free parameters in terms of the shear eigenvalues, we modify the Jedamzik framework as
\begin{equation}
\label{eqn:jedam}
F(M, {\vec\lambda}_{c})
=\int_{M}^{\infty}dM^{\prime}\frac{M^{\prime}}{{\bar\rho}}\frac{dN}{dM^{\prime}}
P(M,M^{\prime}, {\vec\lambda}_{c})\ .
\end{equation}
Here $F(M,{\vec \lambda}_{c})$ is the cumulative probability that the initial shear eigenvalues exceed the 
threshold values of  ${\vec \lambda}_{c}$ on the mass scale of $M$, while 
$P(M,M^{\prime}, {\vec\lambda}_{c})$ is the conditional probability that the shear eigenvalues on the mass scale 
of $M$ exceed the thresholds of ${\vec \lambda}_{c}$ provided that the shear eigenvalues 
on some larger mass scale of $M^{\prime}$ just equal the thresholds. 

In equation (\ref{eqn:jedam}) the cumulative probability, $F(M, {\vec\lambda}_{c})$, can be analytically evaluated 
by integrating the three-point probability density distribution, $p(\lambda_{1},\lambda_{2},\lambda_{3c})$, 
which was derived by \cite{dor70}: 
\begin{equation}
\label{eqn:plambda}
F(M,{\vec \lambda}_{c})= 
\int_{C}\Pi_{i=1}^{n}d^{n} {\lambda}_{i}\,p({\vec\lambda};\sigma), 
\end{equation}
where $n=3,\ 2,\ 1$ for the cases of the 3D, 2D, and 1D EZL, respectively. 
Here the lower-bound $C$ represents the multi-dimensional area over which the integration is performed. 
For the 1D EZL, it is $\{\lambda_{1}\ge\lambda_{2}\ge\lambda_{3}\ge\lambda_{3c}\}$, while for the 2D and 3D 
EZL, they are $\{\lambda_{1}\ge\lambda_{2}\ge\lambda_{2c},
\lambda_{1}\ge\lambda_{2}\ge\lambda_{3}\ge\lambda_{3c}\}$, and 
$\{\lambda_{1}\ge\lambda_{1c},\ \lambda_{1}\ge\lambda_{2}\ge\lambda_{2c},\ 
\lambda_{1}\ge\lambda_{2}\ge\lambda_{3}\ge\lambda_{3c}\}$, respectively

The conditional probability, $P(M,M^{\prime},{\vec\lambda}_{c})$, in equation (\ref{eqn:jedam}) can be also 
obtained under the assumption of Markovian random walks as
\begin{eqnarray}
\label{eqn:pmm1}
P(M,M^{\prime},{\vec\lambda}_{c})&=&P({\vec \lambda}\ge{\vec \lambda}_{c}\vert 
{\vec \lambda^{\prime}}={\vec \lambda}_{c})\,\\
\label{eqn:pmm2}
&=& \int_{C}d^{n}{\lambda_{i}}\, 
p({\vec\lambda}\vert{\vec \lambda}^{\prime}=\lambda_{c})\, 
= \frac{p({\vec\lambda}\ge {\vec\lambda}_{c},\ 
{\vec\lambda}^{\prime}=\lambda_{c})}{p({\vec\lambda}^{\prime}=\lambda_{c})}\nonumber, 
\end{eqnarray}
where ${\vec\lambda}$ and ${\vec\lambda}^{\prime}$ denote the shear eigenvalues on two different mass 
scales $M$ and $M^{\prime}$, respectively. The six-point joint probability density distribution, 
$p({\vec\lambda}, {\vec\lambda}^{\prime})$, has been derived analytically by \cite{des08} and \cite{DS08}. 
For a practical calculation, we discretize equation (\ref{eqn:jedam}) and reexpress it as a matrix product. 
Then we obtain the halo mass function $dN/d\ln M$ as a column vector by converting the matrix equation.

It is  worth noting that our 1D EZL formula is very similar to that of \cite{LS98} in the respect that both 
of the approaches employed the same ellipsoidal collapse condition of $\lambda_{3}=\lambda_{3c}$. 
In the former, however, the halo mass function is automatically normalized while in the latter the halo mass 
function had to be multiplied wrongly by a constant of $12.5$ without taking into account the scale 
dependence of the normalization factor. It is also worth mentioning the key differences between our EZL model 
and that of ref \cite{LL12}. The latter work used the uppler limit on the largest shear eigenvalue, 
$\lambda_{1}\le \lambda_{1c}$ to describe the formation of marginally bound superclusters. 
In the current EZL model for the mass function of bound halos,  there is no such upper limit on the 
shear eigenvalue.  In the next section we determine the best-fit values of the free parameters of our 
EZL formulae by fitting them to  the numerical results from the high-resolution N-body simulations.

\section{Numerical tests}

\subsection{Determination of the barrier heights}\label{sec:height}

\begin{figure}
\centering
\includegraphics[width=14cm]{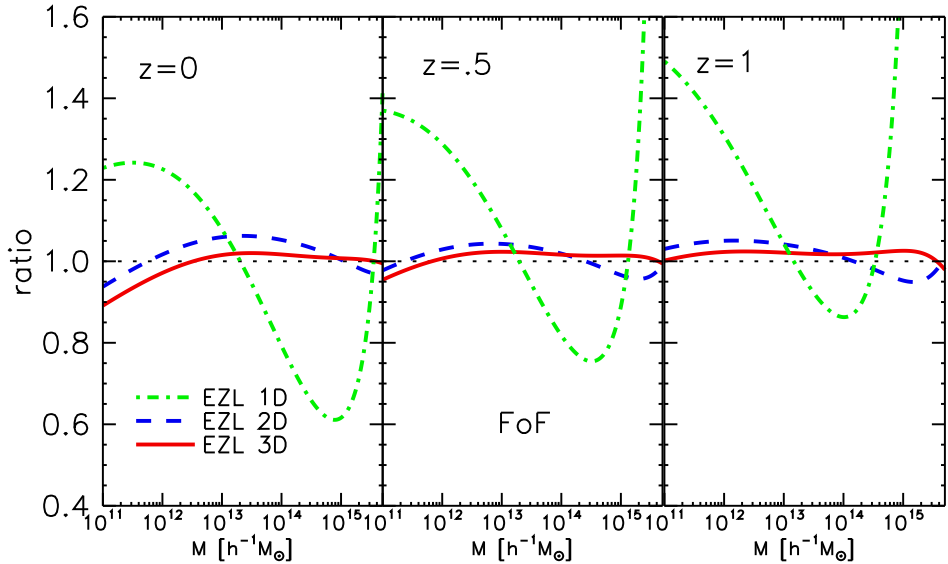}
\caption
{Ratios of the 1D, 2D and 3D EZL mass functions of the FoF halos (green dot dashed, 
blue dashed, and red solid lines, respectively) to the N-body results given in 
ref.~\cite{pph10} as a function of the FoF mass at three different redshifts.}
\label{fig:com_fof}
\end{figure}
\begin{figure}
\centering
\includegraphics[width=14cm]{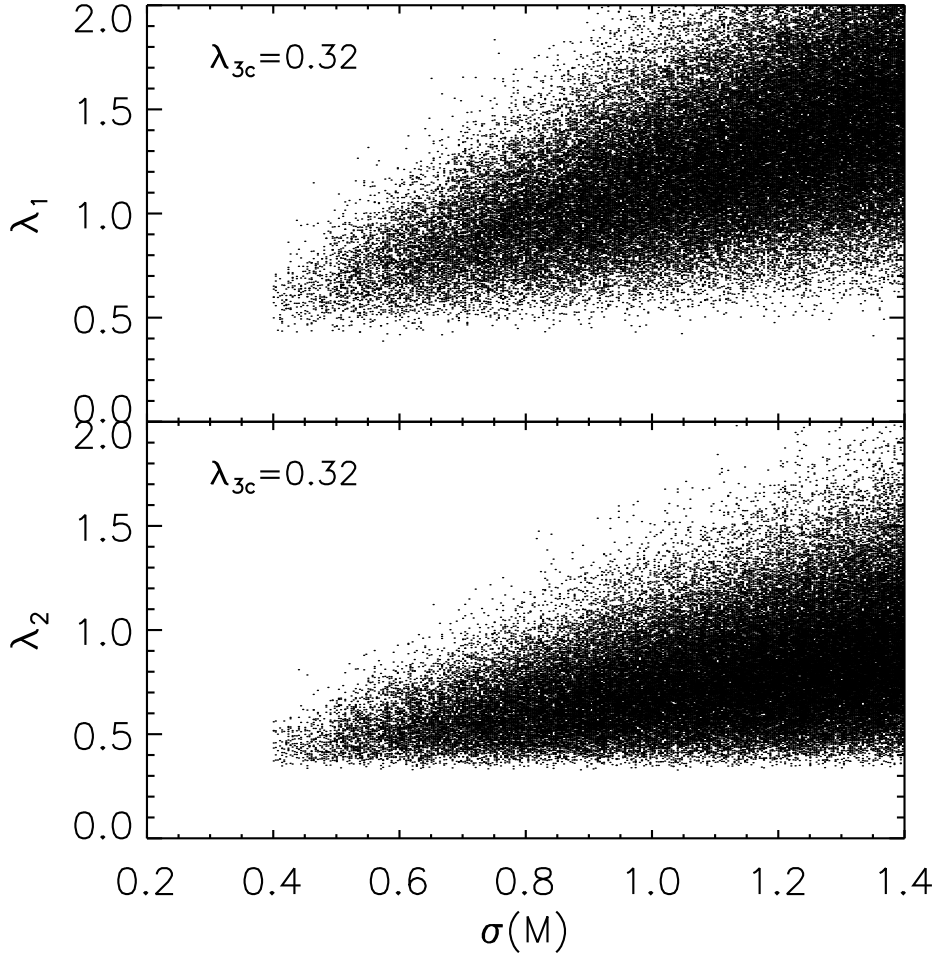}
\caption
{Scatters of the randomly generated shear eigenvalues satisfying the condition of  $\lambda_{3}\ge 0.32$ in the 
$\sigma(M)-\lambda_{1}$ and  $\sigma(M)-\lambda_{2}$ plane in the top and bottom panels, respectively. }
\label{fig:lam12}
\end{figure}
\begin{figure}
\centering
\includegraphics[width=14cm]{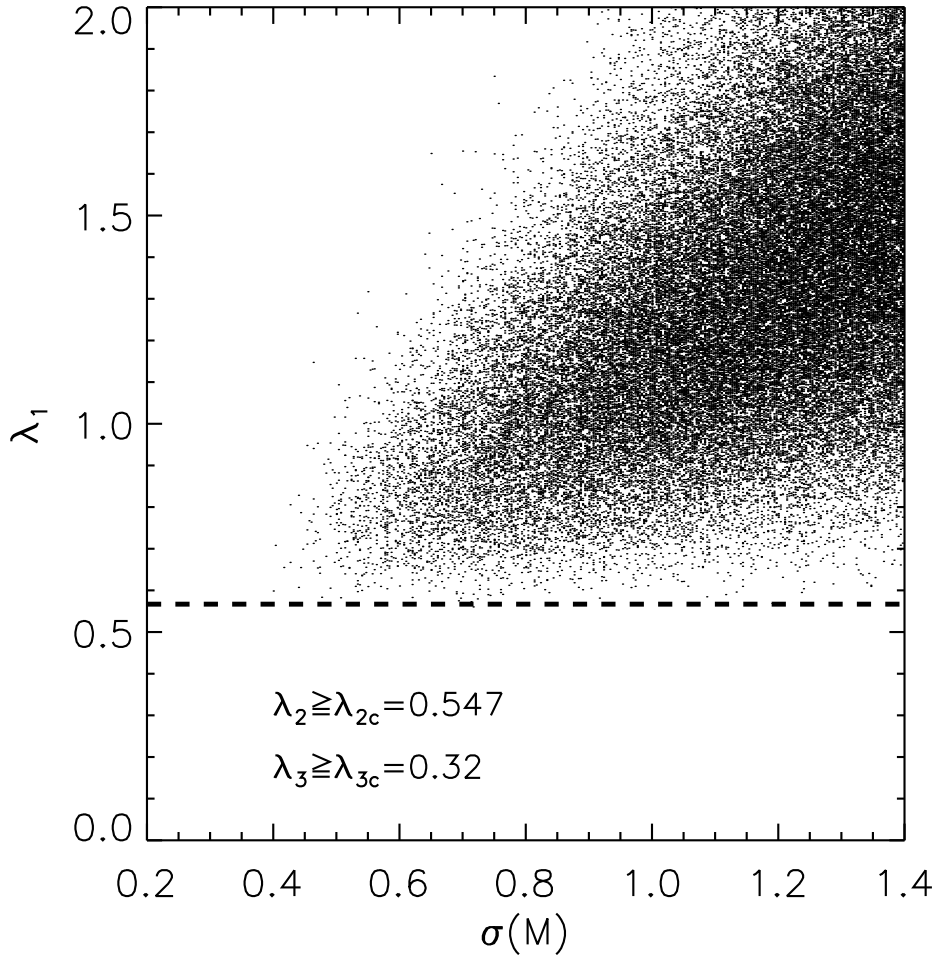}
\caption
{Scatters of the randomly generated shear eigenvalues satisfying the condition of  
$\{\lambda_{2}\ge 0.547,\ \lambda_{3}\ge 0.32\}$ in the $\sigma(M)-\lambda_{1}$ plane. The 
horizontal dashed line corresponds to the 3D EZL threshold of $\lambda_{1c}=0.567$. }
\label{fig:lam1}
\end{figure}
\begin{figure}
\centering
\includegraphics[width=14cm]{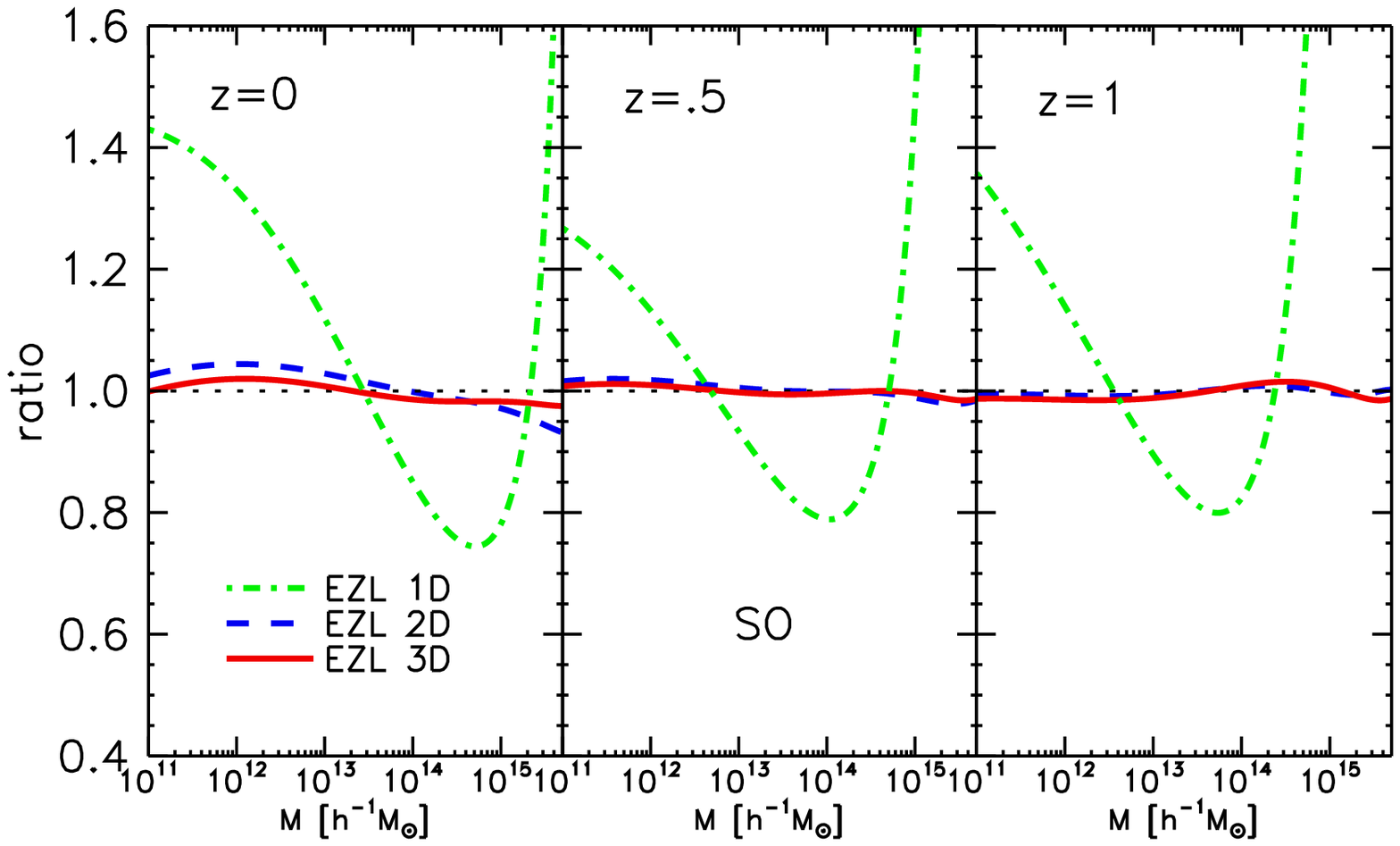}
\caption
{Same as figure \ref{fig:com_fof} but using the Tinker08 formula as the 
numerical results to compare with the EZL mass functions of the SO halos.}
\label{fig:com_so}
\end{figure}
For the empirical determination of the free parameters of  our EZL formula, one has to bear in mind that the 
numerical results from the N-body simulations depend on which halo identification algorithm is used. 
Among the several halo identification algorithms that have been suggested, it is the friends-of-friends (FoF) 
\cite{davis-etal85} and the spherical overdensity (SO) \cite{LC94} algorithms that are most frequently used in the 
literatures.  The former defines a bound halo as the collection of closely packed particles whose separation 
distances are less than some prescribed linking length, while the latter defines a bound halo as the region 
in which the spherically averaged density around its density peak exceeds some critical value.
Given that the mass functions of FoF halos have turned out to be different from those of the SO halos 
\cite[e.g.,][]{tinker-etal08}, we consider both of the FoF and SO cases and determine separately the 
free parameters of the EZL formula for each case. 

For the FoF halo case, we compare our EZL mass function to the numerical results of 
\cite[][hereafter PPH10]{pph10} (see eq. [\ref{eqn:pph}]).  We first conduct a $\chi^{2}$-fitting of the 1D EZL 
formula to the PPH10 at $z=0$ by adjusting only the value of $\lambda_{3c}$ and find the best-fit barrier height 
of the 1D EZL to be $\{\lambda_{3c}=0.41\}$. The top-hat filter is used to calculate the rms density 
fluctuation $\sigma(M)$. Figure \ref{fig:com_fof} shows the ratios of the 1D EZL mass functions  
with this best-fit barrier height to the numerical results of PPH10 as green dot-dashed lines over the wide mass 
range of $10^{11}\le M/(h^{-1}M_{\odot})\le 10^{15}$ at three different redshifts ($z=0,\ 0.5$ and $1$ in the left, 
middle and right panels, respectively).  In each panel, the horizontal dotted line corresponds to unity. As can be 
seen, the agreement between the 1D EZL and the PPH10 is not so good for each case, becoming worse as $z$ 
increases. 

We similarly conduct a $\chi^{2}$ fitting of the 2D and 3D EZL mass functions to the PPH10 by adjusting their 
free parameters, respectively. The best-fit parameters for the 2D and 3D cases  are found to be 
$\{\lambda_{2c}=0.547,\ \lambda_{3c}=0.32\}$ and 
$\{\lambda_{1c}=0.567,\ \lambda_{2c}=0.542,\   \lambda_{3c}=0.32\}$, respectively. 
The best-fit 2D and 3D EZL mass functions are shown as blue dashed and solid red lines, respectively 
in Figure \ref{fig:com_fof}.  As can be seen, the 2D and 3D EZL mass functions show much better 
agreements with the PPH10 than their 1D counterpart at each redshift.   At all redshifts the 3D EZL 
mass function matches the PPH10 best in the whole mass range except for in the range of 
$M\le 10^{12}\,h^{-1}M_{\odot}$  at $z=0$ where the 2D EZL formula works slightly better than its 3D 
counterpart. The ratio of the 2D (3D) EZL mass function to the PPH10 departs from unity by up to $10\%$ 
($7\%$) over the whole mass range at the three redshifts.

Noting that the best-fit value of $\lambda_{2c}$ for the 3D EZL case is very close to that for the 2D EZL case 
and that the 2D EZL formula works almost as well as its 3D counterpart, we investigate how the 2D EZL 
constraints on $\{\lambda_{2},\lambda_{3}\}$ bias the allowed distribution of $\lambda_{1}$.  Generating 
randomly a set of  three shear eigenvalues on the mass scale of $M$ with the help of the Monte-Carlo 
algorithm developed by \cite{CL01}, and selecting only those random points which satisfy
$\lambda_{3}\ge 0.32$, we determine the values of $\lambda_{2}$ and $\lambda_{1}$. 
Then  we repeat the whole process on different mass scales.  For the detailed description of the 
Monte-Carlo algorithm to generate randomly the three shear eigenvalues, see \cite{CL01}.
Figure \ref{fig:lam12} shows the scatter plots of the selected random points in the 
$\sigma(M)-\lambda_{1}$ and $\sigma(M)-\lambda_{2}$ planes in the top and bottom panels, respectively. 
As can be seen,  there are low but non-zero probabilities on all mass scales that $\lambda_{1}$ 
and $\lambda_{2}$ exceed the thresholds of $0.567$ and $0.542$, respectively when $\lambda_{3}\ge 0.32$ is 
satisfied. 

We repeat the same procedure but with the two constraints of $\lambda_{2}\ge 0.547$ and 
$\lambda_{3}\ge 0.32$ to see how biased the values of $\lambda_{1}$ are.  Figure \ref{fig:lam1} 
plots the result in the  $\sigma(M)-\lambda_{1}$ plane. As can be seen, the 3D EZL constraint of $0.567$ is 
automatically satisfied by the cuts on $\lambda_{2}$ and $\lambda_{3}$, which explains why the best-fit value of 
$\lambda_{2c}$ for the 3D EZL case is so close to that for the 2D EZL case as well as why the 2D EZL formula 
works almost as well as the 3D EZL one.

Now, let us turn to the SO halo case for which we use the numerical results of 
\cite[][hereafter, Tinker08]{tinker-etal08} (see eq.[\ref{eqn:tinker}]) to compare our EZL mass functions with. 
Repeating the same $\chi^{2}$-fitting procedure, we determine the best-fit free parameters of the 1D, 2D, and 3D 
EZL formula for the SO-halo case as $\{\lambda_{3c}=0.32\}$, $\{\lambda_{2c}=0.56,\ \lambda_{3c}=0.32\}$
$\{\lambda_{1c}=0.56, \lambda_{2c}=0.557,\ \lambda_{3c}=0.32\}$, respectively.
Figure \ref{fig:com_so} shows the same as figure \ref{fig:com_fof} but for the SO halo case. 
As can be seen,the 2D and 3D EZL mass functions agree excellently with the Tinker08 at all three redshifts. 
The 2D (3D) EZL mass function reaches the accuracy level of maximum $5\%$ ($3\%$) error at all three 
redshifts, which indicates that our EZL formula works well for the FoF and SO cases alike but slightly better for 
the latter case. It is worth emphasizing here that we use the same constant values of ${\vec \lambda_{c}}$ to 
plot the EZL formula at all three redshifts in figures \ref{fig:com_fof}-\ref{fig:com_so}, which implies that the multi-
dimensional barrier heights of our EZL formaula, even though determined empirically at $z=0$, are redshift-
independent. 

\subsection{Direct comparison with the N-body results}
\begin{figure}
\centering
\includegraphics[width=14cm]{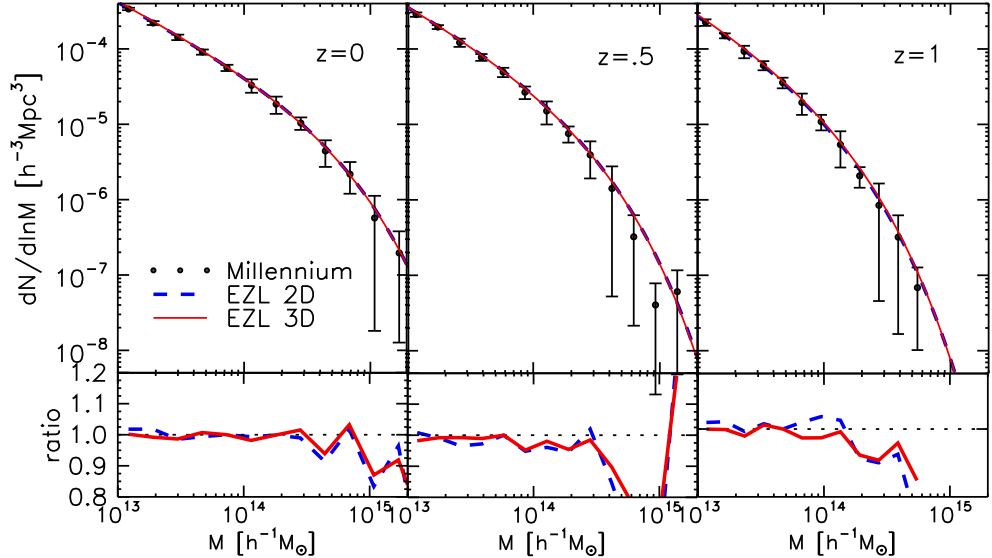}
\caption
{(Top panel): Comparison of the EZL mass functions with the numerical 
results from the Millennium simulations at three different redshifts. 
The parameters of the 2D and 3D EZL models are set at 
$\{\lambda_{2c}=0.547,\ \lambda_{3c}=0.32\}$ and 
$\{\lambda_{1c}=0.567, \lambda_{2c}=0.542,\ \lambda_{3c}=0.32\}$, respectively. 
In each top panel the errors bars are obtained as the standard deviation 
scatter among eight Jack-knife resamples. (Bottom panel): Ratios of the 
EZL mass functions to the N-body results.}
\label{fig:mill}
\end{figure}
\begin{figure}
\centering
\includegraphics[width=14cm]{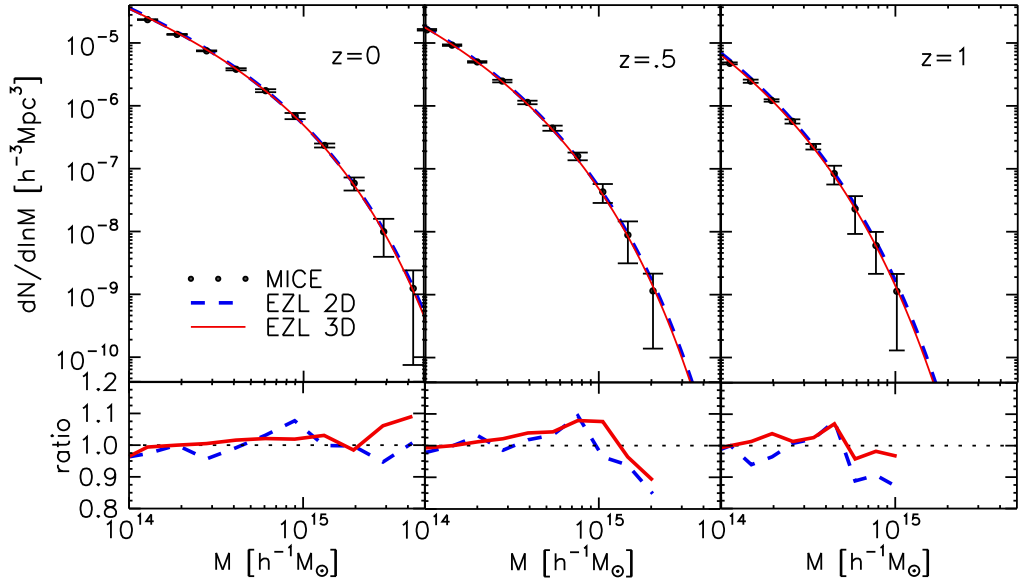}
\caption
{Same as figure \ref{fig:mill} but with the numerical results from the 
MICE simulations. The same best-fit parameters that are  used in figure \ref{fig:mill} are 
also used for the evaluating of the 2D and 3D EZL mass functions}
\label{fig:mice}
\end{figure}

In this section we compare the EZL formula directly with the numerical results 
from two different N-body simulations. First, we use the catalog of the 
FoF halos identified in the Millennium simulations \cite{mill05} for which
the total number of particles ($N_{p}$), individual particle mass, 
linear box size ($L_{p}$) and input cosmological parameters were 
given as 
$N_{p}=10^{10}$, $m_{p}=8.6\times 10^{8}\,h^{-1}M_{\odot}$, 
$L_{p}=0.5\,h^{-1}$Gpc, $\Omega_{m}=0.25,\ \Omega_{\Lambda}=0.75,\ 
h=0.73,\ \sigma_{8}=0.9,\ n_{s}=1.0$, respectively.

Binning the logarithmic masses of the FoF halos from the Millennium 
halo catalog and counting their number densities of those FoFo halos 
belonging to each logarithmic mass-bin per unit volume, we numerically determine, 
$dN/d\ln M$, at $z=0,\ 0.5,\ 1$. The top panel of figure 
\ref{fig:mill} compares the numerical mass functions of the FoF halos 
(dots) with Jack-knife errors from the Millennium simulations with the 2D EZL 
(blue dashed lines) and 3D EZL (red solid lines) mass functions respectively, 
while its bottom panels show the ratios of the EZL mass functions to 
the Millennium results. The Jack-knife errors are calculated as 
one standard deviation scatter among eight Jack-knife resamples \cite{LL12}.
For the plots of the EZL mass functions in each panel, we set the free parameters 
at the best-fit values empirically determined in section \ref{sec:height}: 
 $\{\lambda_{2c}=0.547,\ \lambda_{3c}=0.32\}$ and 
$\{\lambda_{1c}=0.567, \lambda_{2c}=0.542,\ \lambda_{3c}=0.32\}$ for the 2D and 3D cases, 
respectively.  As can be seen, both of the 2D and 3D EZL mass functions agree very well with the 
Millennium results at all redshifts.

We repeat the same calculations but using the catalog of the FoF halos 
identified in the MICE simulations \cite{mice10}: for which
the total number of particles ($N_{p}$), individual particle mass, 
linear box size ($L_{p}$) and input cosmological parameters were 
given as 
$N_{p}=1024^{3}$, $m_{p}=23.42\times 10^{10}\,h^{-1}M_{\odot}$, 
$L_{p}=3.072\,h^{-1}$Gpc, $\Omega_{m}=0.3,\ \Omega_{\Lambda}=0.7,\ 
h=0.7,\ \sigma_{8}=0.8$, respectively. 
Figure \ref{fig:mice} shows the same as figure \ref{fig:mill} but with 
the numerical results from the MICE simulations. As can 
be seen, the EZL mass functions with the same barrier heights show excellent 
agreements with the MICE mass functions at all three redshifts. 
Given that two simulations used two different values of the power spectrum 
amplitude ($\sigma_{8}=0.9$ and $0.8$ for the Millennium and MICE 
simulation, respectively), the results shown in figures 
\ref{fig:mill}-\ref{fig:mice} imply that the multi-dimensional 
barrier height of the EZL mass function should be independent of the background 
cosmology. 

\subsection{Dependence on the filter shape}\label{sec:filter}

\begin{figure}
\centering
\includegraphics[width=13cm]{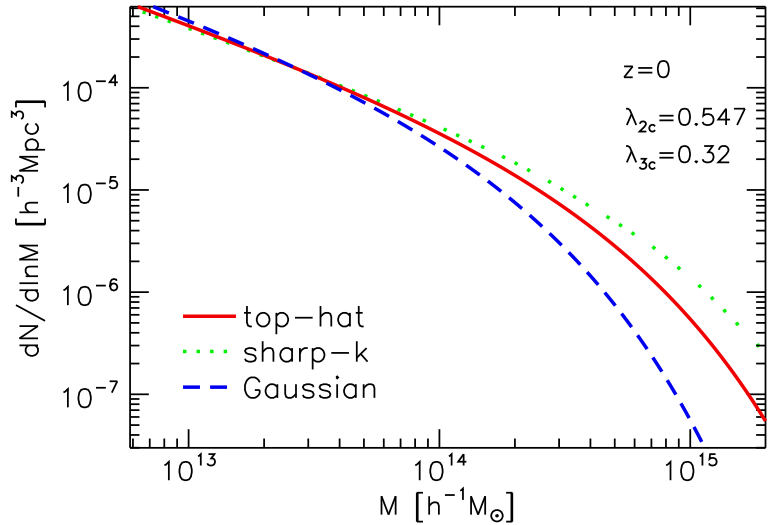}
\caption
{Variation of the 2D EZL mass function with the filter shape. 
The same values of the free parameters 
that was determined in section \ref{sec:height} by using the top-hat filter is applied to all three filter cases.}
\label{fig:filter}
\end{figure}
\begin{figure}
\centering
\includegraphics[width=17cm]{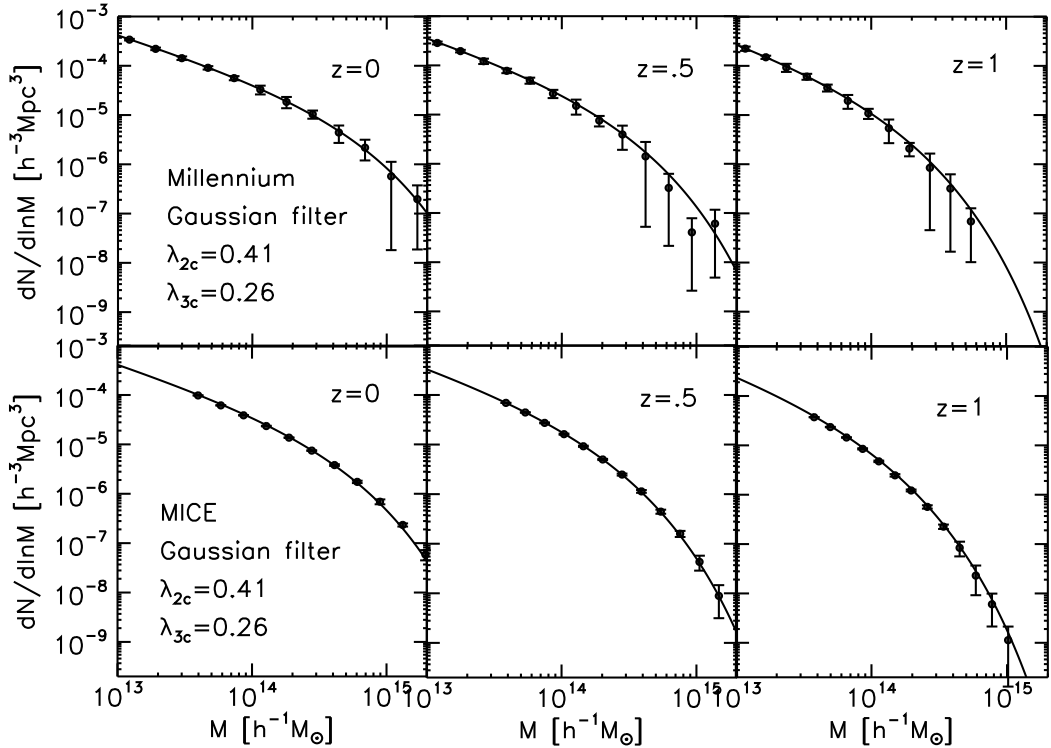}
\caption
{Same as the top panels of figures \ref{fig:mill}-\ref{fig:mice} but for the 
case of the Gaussian filter. }
\label{fig:kfilter}
\end{figure}
\begin{figure}
\centering
\includegraphics[width=14cm]{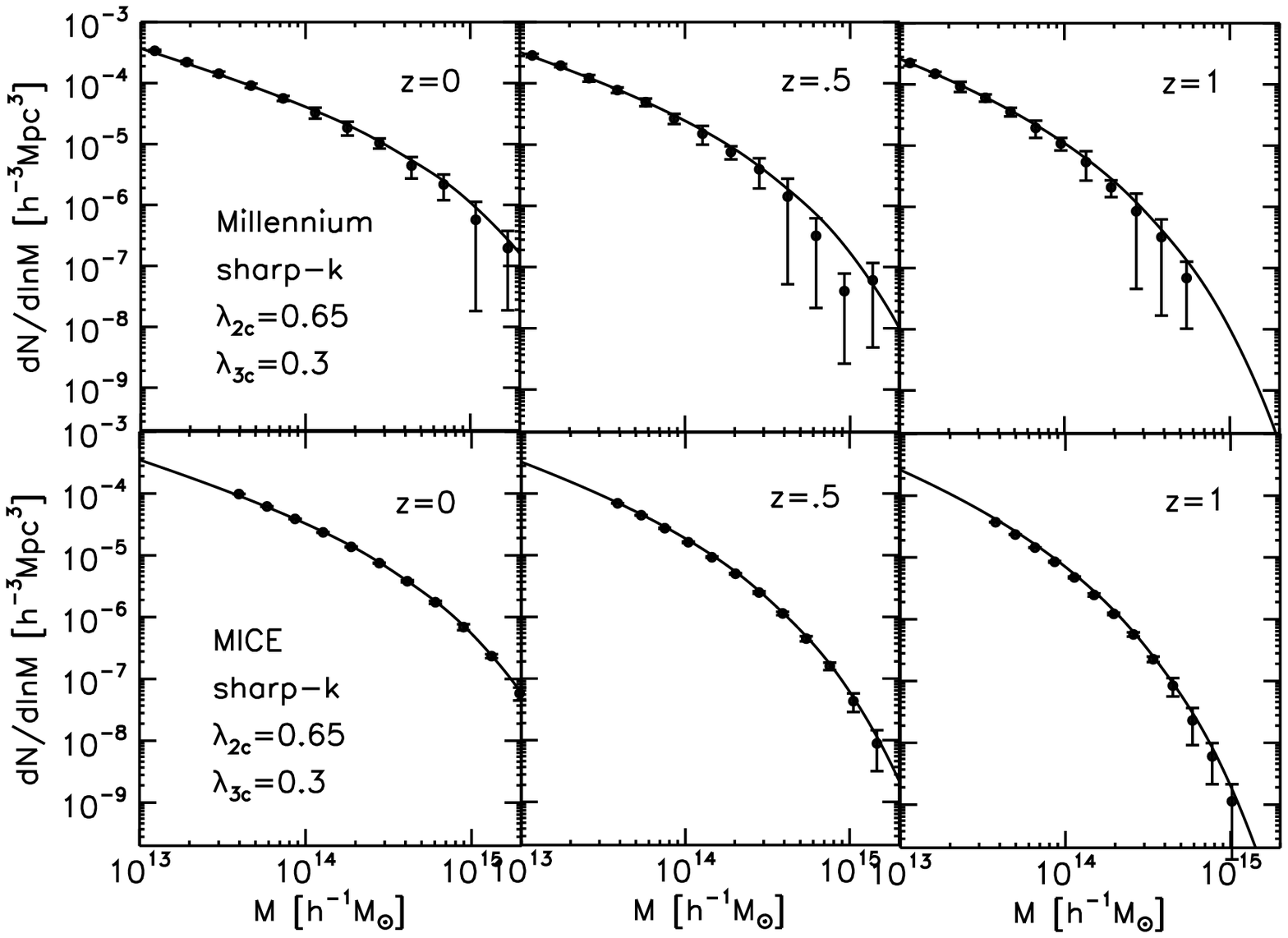}
\caption
{Same as figure \ref{fig:kfilter} but for the case of the sharp-$k$ 
space filter.}
\label{fig:gfilter}
\end{figure}
In the excursion set theory, the mass assignment to a bound halo depends 
on the shape of the filter used to calculate the rms density fluctuation $\sigma(M)$. 
For the results shown in figures \ref{fig:com_fof}-\ref{fig:mice}, 
we have used exclusively the top-hat filter. Using a different filter is likely to 
alter the shape of the EZL mass function and thus accordingly the best-fit 
values of the multi-dimensional barrier heights.
To investigate the dependence of the best-fit values of the free parameters on the 
filter shape, we reevaluate the 2D EZL mass function but 
for the cases of the Gaussian and sharp-k space filter.

Figure \ref{fig:filter} shows the 2D EZL mass functions at $z=0$ for three different filter cases. 
For these plots, we use the same values of $\{\lambda_{2c},\lambda_{3c}\}$ determined 
for the case of the top-hat filter in section \ref{sec:height}. As can be seen, for the Gaussian filter case the 
number densities of the bound halos in the high-mass section ($M\ge 10^{14}\,h^{-1}M_{\odot}$) are 
significantly reduced while those of the low-mass halos increase. 
In contrast, for the sharp-$k$ space filter case the trend becomes reversed: 
more high-mass halos and less low-mass halos than for the top-hat filter case.

To see if the success of the EZL formula is not limited to the top-hat filter case, we repeat the whole fitting 
process described in section \ref{sec:height} to determine the barrier heights of the 2D EZL formula but for the 
Gaussian and the sharp-$k$ space filter cases. Figures \ref{fig:kfilter} and \ref{fig:gfilter} show the 2D EZL mass 
functions (solid line) with the best-fit parameters at three different redshifts for the Gaussian and sharp-$k$ 
space filter cases, respectively. In each figure,  the N-body results from the Millennium and the MICE 
simulations are also shown as dots (in the top and bottom panels, respectively). 
The best-fit parameters for both of the cases are different 
from those for the top-hat filter case. The higher value of $\lambda_{2c}$ 
and lower value of $\lambda_{3c}$ for the sharp-$k$ filter case, 
while the lower values of $\lambda_{2c}$ and $\lambda_{3c}$ for the 
Gaussian filter case. Anyway, even for the Gaussian and sharp-$k$ space 
filter case the 2D EZL mass functions with the constant values of the parameters at all redshifts are 
in excellent agreements with the N-body results.

\section{Discussion}\label{sec:limit}

Despite its excellent agreements with various N-body results,  it is not a physical model but only a 
phenomenological fitting formula for the halo mass function since the best-fit values of  its 
parameters are not related to the physics of real halo formation process. 
A critical reader might then ask what the advantage of using such a more complicated matrix equation 
as equation (\ref{eqn:jedam}) for the evaluation of the halo mass function rather than using those simpler 
formulae suggested in the previous works. 
The merit of our EZL model is that it automatically satisfies the normalization constraint, as mentioned 
in \ref{sec:ezl}, providing accurate fits with fewer fitting parameters than the previous models.
Note that although the previous formulae of ref.~\cite{tinker-etal08} and ref.~\cite{pph10} have four fitting 
parameters including the normalization factor, their accuracy levels are not higher than our 2D EZL model which 
has only two parameters. Moreover, the empirically determined values of the free parameters
of our EZL model have turned out to be independent of redshift and background cosmology. In our companion 
paper (S. Lim \& J. Lee 2012 in preparation), we have found that the same best-fit values of the parameters
work even for a cosmology with primordial non-Gaussianity.

At any rate, to develop a physical model for the halo mass function within the framework of the Zel'dovich 
approximation, it will be evidently necessary  to describe the growth of the initial shear eigenvalues by  
three dimensional {\it non-Markovian} random walks. It has been recently proved by ref.~\cite{PLS12} that 
neglecting the correlations among random walks in the excursion set theory can significantly flaw
analytic evaluation of the halo mass function. As an example, ref.~\cite{PLS12}  demonstrated that in the 
asymptotic limit of completely correlated random walks the excursion set theory in fact recovers 
the original PS mass function without the normalization factor of $2$. 
Regarding the normalization issue that the overall amplitude of the excursion set mass function for the 
non-Markovian case is substantially lower than the numerical results, ref.~\cite{PS12} has argued that it 
should imply the limitation of the excursion set approach itself: Due to the wrong assumption that 
the differential mass function $dN/dM$ on the right hand side of equation (\ref{eqn:frame}) can be simply 
related to the statistics of randomly placed cells on which the calculation of the differential volume fraction 
$f(\sigma)$ in the left-hand side of equation (\ref{eqn:frame}) is based, 
the excursion set mass function for the realistic non-Markovian case underestimates the abundance of bound 
halos. 

The recent work of \cite{MS12} may provide a clue  to how to incorporate the non-Markovian nature of random 
walks of  shear eigenvalues into the Zel'dovich model. Ref.~\cite{MS12} have derived  a very simple and 
accurate analytic model for the distribution of the non-Markovian random walks that first cross a given collapse 
barrier of arbitrary shape, and claimed that their model is valid even for the case that the collapse barrier is not a 
function of linear density contrast but of other initial quantities. Our future work is in the direction of finding a 
relation of the shear eigenvalues to the physics of true halo formation by extending the work of \cite{MS12} 
to the case of three dimensional non-Markovian random walks of shear eigenvalues \cite[see also][]{SCS12}.

\acknowledgments 

We thank an anonymous referee who helped us improve significantly the original manuscript.
We acknowledge the use of data from the Millennium and the MICE 
simulations that are publicly available at http://www.millennium.com, \\
and http://www.ice.cat/mice, respectively. 
The Millennium Simulation analyzed in this paper was carried out by the 
Virgo Supercomputing Consortium at the Computing Center of the Max-Planck 
Society in Garching, Germany. The computer code which evaluates the EZL halo 
mass functions  will be provided upon request.
This work was supported by the National Research Foundation of Korea (NRF) 
grant funded by the Korea government (MEST, No.2012-0004196) and partially 
by the research grant from the National Research Foundation of Korea to the 
Center for Galaxy Evolution Research  (NO. 2010-0027910).


\begin{thebibliography}{99}

\bibitem{peebles02}
P.~J.~E.~Peebles, {\it From Precision Cosmology to Accurate Cosmology}, 
2008 {\it proceedings of the Moriond Conference on the Cosmological 
Models} [astro-ph/0208037]

\bibitem{PS74} Press W H and Schechter P, 
{\it Formation of galaxies and clusters of galaxies by self-similar 
gravitational condensation}, 1974 {\it Astrophys. J.} {\bf 187} 425

\bibitem{bond-etal91} Bond J R, Cole S, Efstathiou G and Kaiser N,
{\it Excursion set mass functions for hierarchical Gaussian fluctuations}, 
1991 {\it Astrophys. J.} {\bf 379} 440 

\bibitem{PH90} Peacock J A and Heavens A F, 
{\it Alternatives to the Press-Schechter cosmological mass function}, 
1990 {\it Mont. Not. Roy. Astron. } {\bf 243} 133

\bibitem{jedam95}
Jedamzik K, {\it The Cloud-in-Cloud Problem in the Press-Schechter 
Formalism of Hierarchical Structure Formation }, 
1995 {\it Astrophys. J.} {\bf 448} 1 

\bibitem{yano-etal96} Yano T, Nagashima M, and Gouda N,
{\it Limitations of the Press-Schechter Formalism}, 
1996 {\it Astrophys. J.} {\bf 466} 1 

\bibitem{audit-etal97} Audit E, Teyssier R and Alimi J M,  
{\it Non-linear dynamics and mass function of cosmic structures. I. 
Analytical results}, 1997 {\it Astron. \& Astrophys.} {\bf 325} 439

\bibitem{monaco97a} Monaco P., {\it A Lagrangian Dynamical Theory for the 
Mass Function of Cosmic Structures - I. Dynamics}, 
1997 {\it Mont. Not. Roy. Astron. } {\bf 287} 753

\bibitem{monaco97b} Monaco P., {\it A Lagrangian dynamical theory 
for the mass function of cosmic structures - II. Statistics}, 
1997 {\it Mont. Not. Roy. Astron. } {\bf 290} 439 

\bibitem{CL01}  
Chiueh T and Lee J, {\it On the Nonspherical Nature of Halo Formation}, 
2001 {\it Astrophys. J.} {\bf 555} 83

\bibitem{SMT01} Sheth R K, Mo H J and Tormen G, 
{\it Ellipsoidal collapse and an improved model for the number and 
spatial distribution of dark matter haloes}, 2001 
{\it Mont. Not. Roy. Astron. Soc.} {\bf 323} 

\bibitem{MR10a}  Maggiore M and Riotto A, 
{\it The halo mass function from excursion set theory. I. 
Gaussian fluctuations with non-Markovian dependence on the smoothing scale}, 
2010 {\it Astrophys. J.} {\bf 711} 907 

\bibitem{MR10b} Maggiore M and Riotto A, 
{\it The Halo mass function from excursion set theory. II. The diffusing 
barrier} 2010 {\it Astrophys. J.} {\bf 717} 515 [arXiv:0903.1250]

\bibitem{MR10c} Maggiore M and Riotto A, 
{\it The halo mass function from excursion set theory. III. 
non-Gaussian fluctuations}, 2010 {\it Astrophys. J.} {\bf 717} 526

\bibitem{CA11a} Corasaniti P S and Achitouv I, 
{\it Toward a universal formulation of the halo mass function},
2011, {\it Phys. Rev. Lett.} {\bf 106} 241302

\bibitem{CA11b} Corasaniti P S and Achitouv I, 
{\it Excursion set halo mass function and bias in a stochastic 
barrier model of ellipsoidal collapse}, 2011 {\it Phys. Rev. D} 
{\bf 84} 023009

\bibitem{PLS12} Paranjape A, Lam T Y and Sheth R K
{\it Halo abundances and counts-in-cells: the excursion set approach with correlated steps}, 
2012 {\it Mont. Not. Roy. Astron. Soc.} {\bf 420} 1429

\bibitem{MS12} 
Musso M and Sheth R K, 
{\it One step beyond: the excursion set approach with correlated steps}, 
2012 {\it Mont. Not. Roy. Astron. Soc.} {\bf 423} L102

\bibitem{PS12} Paranjape A and Sheth R K
{\it Peaks theory and the excursion set approach}, 
2012  [arXiv:1206.3506]

\bibitem{LS98} Lee J and Shandarin S F, {\it The Cosmological Mass 
Function in the Zel'dovich Approximation}, 1998 {\it Astrophys. J.} 
{\bf 500} 4
 
\bibitem{ST99} Sheth R K and Tormen G, 
{\it Large-scale bias and the peak background split}, 1999 
{\it Mont. Not. Roy. Astron.} {\bf 308} 119

\bibitem{reed-etal03} 
Reed D, Gardner J, Quinn T, Stadel J, Fardal M, Lake G, and Governato F, 
{\it Evolution of the mass function of dark matter haloes}, 2003 
{\it Mont. Not. Roy. Astron. Soc.}  {\bf 346} 565

\bibitem{jenkins-etal01}
Jenkins A et al.,{\it The mass function of dark matter haloes}, 2001 
{\it Mont. Not. Roy. Astron. Soc.}  {\bf 321} 372

\bibitem{warren-etal06}
Warren M S, Abazajian K, Holz D E, and Teodoro L, 
{\it Precision Determination of the Mass Function of Dark Matter Halos}, 
2006 {\it Astrophys. J.} {\bf 646} 881

\bibitem{tinker-etal08} Tinker J L et al., {\it Toward a halo mass 
function for precision cosmology: The limits of universality}, 
2008 {\it Astrophys. J.} {\bf 688} 709

\bibitem{mice10}
Crocce M, Fosalba P, Castander F J, and Gaztanaga E, 
{\it Simulating the Universe with MICE: the abundance of massive clusters}, 
2010 {\it Mont. Not. Roy. Astron. Soc.}  {\bf 403} 1353

\bibitem{pph10} Pillepich A, Porciani C and Hahn O, {\it Halo 
mass function and scale-dependent bias from N-body simulations with 
non-Gaussian initial conditions}, 2010 {\it Mont. Not. Roy. Astron. Soc.}  
{\bf 402} 191

\bibitem{lee12} Lee J, {\it The relative abundance of isolated cluster 
as a probe of Dark energy}, 2012 {\it Astrophys. J.} {\bf 752} 40

\bibitem{zel70} Zel'dovich Y B, 
{\it Gravitational instability: An approximate theory for large density 
perturbations.}, 1970 {\it Astron. \& Astrophys.} {\bf 5} 84

\bibitem{wmap7} Komatsu E et al., {\it Seven-year Wilkinson 
microwave anisotropy probe (WMAP) observations: Cosmological 
interpretation}, 2011 {\it Astrophys. J. Supp.} {\bf 192} 18

\bibitem{camb} Lewis A and Bridle S, {\it Cosmological parameters 
from CMB and other data: A Monte Carlo approach}, 2005
{\it Phys. Rev. D.} {\bf 66} 103511

\bibitem{eke-etal96} 
Eke V R, Cole S, and Frenk C S, 
{\it Cluster evolution as a diagnostic for Omega}, 1996 
{\it Mont. Not. Roy. Astron. Soc.}  {\bf 282} 263

\bibitem{davis-etal85}  
Davis M, Efstathiou G, Frenk C S, and White S D M., 
{\it The evolution of large-scale structure in a universe dominated by 
cold dark matter}, 1985 {\it Astrophys. J.} {\bf 292} 371

\bibitem{LC94} Lacey C and Cole S, 
{\it Merger Rates in Hierarchical Models of Galaxy Formation - 
Part Two - Comparison with N-Body Simulations}, 
1994 {\it Mont. Not. Roy. Astron.} {\bf 271} 676 

\bibitem{wmap5} Dunkley J et al., 
{\it Five-Year Wilkinson Microwave Anisotropy Probe Observations: 
Likelihoods and Parameters from the WMAP Data}, 
2009 {\it Astrophys. J.} {\bf 180} 306

\bibitem{BM96} Bond J R and Myers S T, 
{\it The Peak-Patch Picture of Cosmic Catalogs. I. Algorithms}, 
1996 {\it Astrophys. J.} {\bf 103} 1 

\bibitem{porciani-etal02} 
Porciani C, Dekel A, and Hoffman Y, 
{\it Testing tidal-torque theory - II. Alignment of inertia and shear and the characteristics of protohaloes}, 2002 
{\it Mont. Not. Roy. Astron. Soc.}  {\bf 332} 339

\bibitem{LL12} Lim S and Lee J, 
{\it The excursion set mass function of superclusters}, [arXiv:1201.1382]

\bibitem{dor70} Doroshkevich A G, {\it Spatial structure of perturbations 
and origin of galactic rotation in fluctuation theory}, 
1970 {\it Astrophyzika} {\bf 3} 175

\bibitem{des08} Desjacques V, {\it Environmental dependence in the
ellipsoidal collapse model}, 2008 {\it Mont. Not. Roy. Astron.} 
{\bf 388} 638 

\bibitem{DS08} Desjacques V and Smith R E, 
{\it Statistical properties of the linear tidal shear}, 
2008 {\it Phys. Rev. D.} {\bf 66} 103511

\bibitem{mill05} Springel V et al., 
{\it Simulations of the formation, evolution and clustering of galaxies 
and quasars}, 2005 Nature {\bf 435} 629 

\bibitem{SCS12} 
Sheth R  K, Chuen C K, and Scoccimarro R, 
{\it Non-local Lagrangian bias}, 2012  [arXiv:1207.7117] 

\end{thebibliography}
\end{document}